\documentclass{article}
\usepackage{spconf,amsmath,graphicx,hyperref}
\usepackage{booktabs} 
\usepackage{makecell}
\usepackage{textcomp}
\usepackage{xcolor}

\title{Training Dynamics-Aware Multi-Factor Curriculum Learning for Target Speaker Extraction}
\name{
  Yun Liu$^{1,2}$,
  Xuechen Liu$^{1}$,
  Xiaoxiao Miao$^{3}$,
  Junichi Yamagishi$^{1,2}$
}

\address{
  $^{1}$National Institute of Informatics, Tokyo, Japan \\
  $^{2}$Sokendai, Kanagawa, Japan \\
  $^{3}$Division of Natural and Applied Sciences, Duke Kunshan University, Kunshan, China   \\
  \texttt{\{yunliu, xuecliu, jyamagis\}@nii.ac.jp}
}

\begin{document}
\ninept
\maketitle
\begin{abstract}
Target speaker extraction (TSE) aims to isolate a specific speaker's voice from multi-speaker mixtures. Despite strong benchmark results, real-world performance often degrades due to different interacting factors. Previous curriculum learning approaches for TSE typically address these factors separately, failing to capture their complex interactions and relying on predefined difficulty factors that may not align with actual model learning behavior.
To address this challenge, we first propose a multi-factor curriculum learning strategy that jointly schedules SNR thresholds, speaker counts, overlap ratios, and synthetic/real proportions, enabling progressive learning from simple to complex scenarios. However, determining optimal scheduling without predefined assumptions remains challenging. We therefore introduce TSE-Datamap, a visualization framework that grounds curriculum design in observed training dynamics by tracking confidence and variability across training epochs. Our analysis reveals three characteristic data regions: (i) easy-to-learn examples where models consistently perform well, (ii) ambiguous examples where models oscillate between alternative predictions, and (iii) hard-to-learn examples where models persistently struggle. Guided by these data-driven insights, our methods improve extraction results over random sampling, with particularly strong gains in challenging multi-speaker scenarios.

\end{abstract}
\begin{keywords}
Target speaker extraction, Curriculum learning, Data selection, Training dynamics.
\end{keywords}
\section{Introduction}
Target speaker extraction (TSE) aims to isolate a target voice from mixtures with other speakers and noise \cite{2023Neural, 2019SpeakerBeam}. Conventional TSE training schemes employ uniform random sampling across all training data \cite{Xu2020SpEx,Tao2025Audio-Visual}, treating examples equally regardless of their learning difficulty. However, established research has demonstrated that models exhibit varying learning difficulty across different samples, motivating the adoption of curriculum learning (CL) \cite{Wang2021A} that progressively introduce training examples from easy to hard based on difficulty factors. Well-established difficulty factors include signal-to-noise ratio (SNR) \cite{Gao2016SNR-Based, Tu2020A}, the number of interfering speakers \cite{Wang2024Enhancing, Chen2020Continuous}, and temporal overlap \cite{Lin2021Sparsely, Pan2021USEV}. With the increasing use of synthetic data for training augmentation, recent studies have also identified the nature of interfering speakers (real versus synthetic) as another emerging difficulty \cite{Liu2024Improving, Liu2024Libri2Vox}. In real-world scenarios, these factors do not operate independently, and their interactions create challenges that cannot be anticipated by analyzing each factor in isolation.

Previous CL approaches for TSE typically address these factors separately \cite{Liu2024Target}, progressively increasing difficulty along a single dimension. However, such single-factor curricula fail to capture complex factor interactions and rely on predefined difficulty metrics that may not align with how models actually perceive task difficulty during training \cite{Wang2021A,Xu2022Confidence-Aware}. This mismatch can result in ineffective curriculum scheduling, where examples considered ``easy" by predefined metrics may actually be challenging for the model to learn.

To address these limitations, we propose a multi-factor curriculum learning strategy that jointly schedules \textbf{SNR thresholds, speaker counts, overlap ratios, and synthetic/real proportions}, enabling models to learn progressively from simple to complex scenarios. The key challenge here lies in determining the optimal scheduling of these multiple factors without relying on predefined assumptions about task difficulty. Rather than using predetermined setups, we extend our curriculum design in observed training dynamics and propose \textbf{\emph{TSE-Datamap}}, a data selection and visualization framework that maps training examples according to how models actually learn them over time \cite{Swayamdipta2020Dataset}. By tracking basic statistics across epochs for each example, we construct a 2-dimensional representational space with distinct regions, representing different level of learning difficulty for the model.
Our analysis reveals that models achieve a more efficient optimization process when first exposed to easy-to-learn samples, which provide clear separation cues and establish reliable decision boundaries before tackling more complex cases. This approach ensures that curriculum scheduling aligns with actual model learning behavior rather than predetermined assumptions about task difficulty.

This paper makes two main contributions: 1) We propose a multi-factor CL strategy for TSE that jointly schedules multiple complexity factors, enabling progressive learning from simple to complex scenarios; 2) We introduce TSE-Datamap, grounding the proposed CL design in observed training dynamics rather than predefined difficulty metrics. Experiments show that our approach consistently outperforms single-factor curricula, with substantial performance gain in complex multi-speaker scenarios.

\section{Method}

\subsection{Target Speaker Extraction}

Target speaker extraction aims to recover the speech of a target speaker from a mixture containing multiple concurrent speakers and background noise. Given a single-channel mixture:
\begin{equation}
y(t) = s_{\text{tar}}(t) + \sum_{k=1}^{K} u_k(t) + n(t)
\label{eq:tse_def}
\end{equation}
where $s_{\text{tar}}(t)$ is the target speaker's speech, $u_k(t)$ is $k$-th interfering speaker of $K$ speakers, and $n(t)$ represents background noise. A TSE model $f_\theta(\cdot)$ takes the mixture $y$ and a reference utterance $c$ from the target speaker as inputs, producing an estimate $\hat{s}_{\text{tar}} = f_\theta(y, c)$. The model is trained to minimize a regression loss between the estimated and ground-truth target speech.

\subsection{Multi-Factor Complexity in TSE}
As previously mentioned, the difficulty of performing TSE on a multi-speaker mixture depends on multiple factors that interact in complex manner. This study focuses on four key factors and explores their impact on speech extraction performance. Below describes them in detail.

\textbf{Signal-to-Noise Ratio (SNR).} Given the mixture formulation in Eq. \ref{eq:tse_def}, the SNR is defined as $\text{SNR} = 10\log_{10}\frac{\|s_{\text{tar}}(t)\|^2}{\|\sum_{k=1}^K u_k(t) + n(t)\|^2}$, where lower SNR values indicate stronger interference relative to the target signal. When SNR approaches or falls below 0 dB, the interference power dominates the target speech, resulting in severe masking effects that significantly impair the model's ability to distinguish target speech characteristics from background interference.

\textbf{Number of Interfering Speakers.} Conventional TSE involves only single interfering speaker \cite{2019SpeakerBeam, Xu2020SpEx}, where the model distinguishes between two speakers in the mixture. With multiple interfering speakers, the model must isolate the target from complex spectral interactions created by overlapping speech patterns. The acoustic confusion grows non-linearly as the number of concurrent speakers $K$ in Eq. \ref{eq:tse_def} increases, since each new speaker introduces additional spectral overlap and potential masking effects across different frequency bands \cite{Wang2024Enhancing, Chen2020Continuous}.

\textbf{Temporal Overlap Ratio.} The temporal overlap ratio quantifies the proportion of time when target and interfering speakers produce speech simultaneously, causing unpredictable impact on the underlying acoustic and speaker cues. Low overlap scenarios provide abundant clean target that serve as clear acoustic references. On the other hand, high overlap ratios may transform the task into concurrent source separation without frequent speaker transitions, a configuration that modern TSE models can handle with relative stability. Intermediate overlap ratios present the greatest challenge, as rapid alternations between target and interfering speakers demand continuous speaker tracking and robust handling of frequent acoustic transitions \cite{Taherian2021Time-Domain, Chen2020Continuous}. This is also the case for modern TSE systems, as established in earlier works \cite{Li2023Audio-Visual,Pan2021USEV}. In this work, the ratio ranges from 0 to 1, where 0 indicates full overlap and 1 represents no overlap between the speakers in the mixture.

\textbf{Synthetic and real interfering speakers.} The use of synthetic interfering speakers introduces variations that may not exist in real recordings. Synthetic speech data can occupy unusual regions of the acoustic space, creating challenges that differ from those in real speech interference. While real interfering speakers provide authentic acoustic patterns, synthetic data can help models learn to handle edge cases and prevents overfitting to specific speaker characteristics \cite{Wang2024Speech,Liu2024Improving}.

\subsection{TSE-Datamap}
\label{datamap2}
This section introduces TSE-Datamap, the proposed data selection and visualization framework. Different from conventional CL, which employs hand-crafted rules and ignores feedback from the model during training, the training examples in the proposed method are placed according to how the model learns it over time. By tracking the model behavior, TSE-Datamap reveals three regions (easy, ambiguous, hard) that directly guide our sampling schedule.

\subsubsection{Fundamental Terms}
For each training example $i$, we use the training loss as a metric, denoted as $M_i$. We then calculate two key statistics over E training epochs to analyze the overall training dynamics: The \textbf{confidence} $\mu_i = \frac{1}{E} \sum_{e=1}^{E} M_i^{(e)}$ measures the mean metric across epochs, where $M_i^{(e)}$ is the metric for example $i$ at epoch $e$. The \textbf{variability} $\sigma_i = \sqrt{\frac{1}{E} \sum_{e=1}^{E} (M_i^{(e)} - \mu_i)^2}$ captures the consistency of predictions through the standard deviation (std). 

We use an SNR-based loss that directly maximizes signal quality. Within each epoch, $M_i(s_{\text{tar},i}, \hat{s}_{\text{tar},i}) = -10 \log_{10} \frac{\|s_{\text{tar},i}\|^2}{\|s_{\text{tar},i} - \hat{s}_{\text{tar},i}\|^2}$, where $s_{\text{tar},i}$ and $\hat{s}_{\text{tar},i}$ are the target and predicted speech respectively for each training example $i$.
The confidence and variability create a 2-dimensional datamap, where each training example occupies a specific position based on its training dynamics. which revealing three distinct learning behaviors:
\begin{itemize}
\item \textbf{Easy-to-learn}: These have \textbf{high confidence} and \textbf{low variability}, and typically include clear, high-SNR speech with minimal interference.
\item \textbf{Ambiguous}: Characterized by \textbf{high variability}, these show rather unstable predictions as the model oscillates between different hypotheses. They samples are rich in discriminative information, forcing the model to learn robust boundaries and improving generalization. In TSE, this occurs with moderate overlap or acoustically similar speakers.
\item \textbf{Hard-to-learn}: These have \textbf{low confidence} and \textbf{low variability}, indicating rather imprecise predictions. These occur often due to extremely adverse conditions like very low SNR where discriminative cues are minimal.
\end{itemize}

\subsubsection{Construction and visualization}
During training, we generate data by uniformly sampling four factors for each training example: SNR from $\{0, 5, 10, 15\}$ dB, overlap ratio from $\{0, 0.2, 0.4\}$ (where 0 indicates full overlap), number of interfering speakers from $\{1, 2, 3\}$, and interference type from $\{\text{real}, \text{syn}, \text{real/syn}\}$, where “\text{real/syn}” means that for multi-speaker mixtures, each interfering speaker is randomly selected from either the real or synthetic pool. 

\begin{figure}[t]
  \centering
  \includegraphics[width=\linewidth]{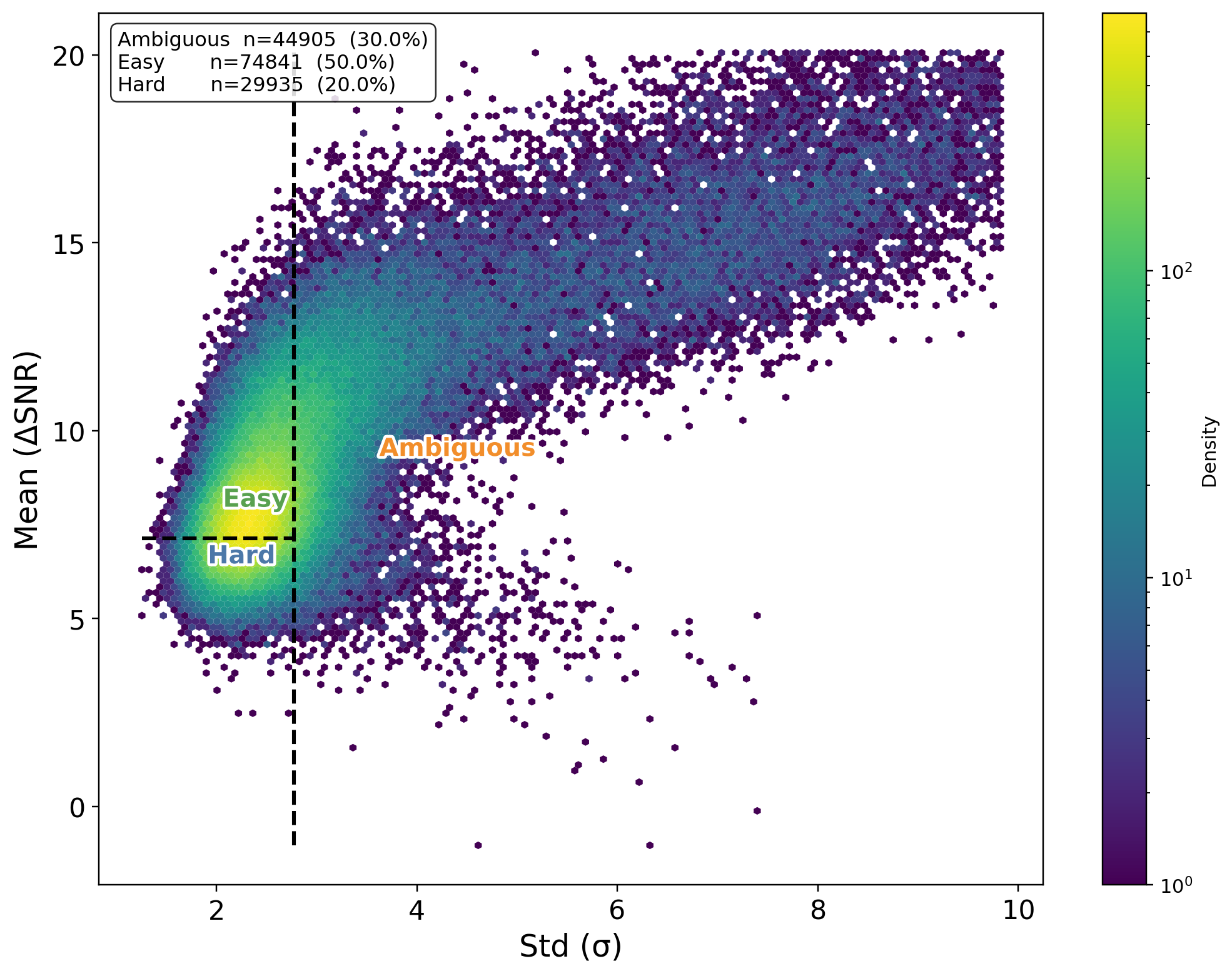}
  \caption{TSE-Datamap computed. Each point denotes a sample, with $x$-axis the standard deviation of $\Delta\mathrm{SNR}$ and $y$-axis the mean $\Delta\mathrm{SNR}$. Number of training epochs in this study is 50.}
\label{fig:datamap_pro}
\end{figure}

To construct the TSE-Datamap visualization shown in Fig.~\ref{fig:datamap_pro}, we track the training dynamics of each example across $E=50$ epochs. Specifically, for each training sample $i$, we record the improved SNR loss($\Delta\mathcal{L}_{\text{SNR}}$) at every epoch $e$, computed as the difference between output and input SNR. We then calculate confidence across different epochs according to definitions in Section 2.3,  discarding the first epoch to mitigate initial training instability. In this study, for simplicity, we use $\Delta\mathcal{L}_{\text{SNR}}$ because our training objective is SNR-based,  noting that this procedure is essentially metric-agnostic, and any measure that reflects extraction quality (e.g., signal-to-distortion ratio (SDR), STOI, PESQ, DNSMOS \cite{Reddy2020Dnsmos}) can be substituted.

The $\Delta\mathcal{L}_{\text{SNR}}$ of each training example across the training epochs is then plotted as a point in this 2D space, with variability on the x-axis and confidence on the y-axis. Based on the resulting distribution, we categorize examples into three regions: we first select 30\% of samples with highest variability as \textbf{Ambiguous}, then from the remaining 70\%, we classify the top 50\% with highest confidence as \textbf{Easy-to-learn} and the bottom 20\% as \textbf{Hard-to-learn}. This data-driven categorization, combined with the tracked metadata of the four complexity factors, enables analysis of how different factor combinations distribute across the map and informs our adaptive CL strategy that responds to actual model behavior rather than predetermined difficulty assumptions.

\section{Dataset and Training Configuration}

\subsection{Dataset Composition}
We acquire \emph{Libri2Vox} as the main dataset\cite{Liu2024Libri2Vox}, which is a mixture of LibriTTS \cite{zen2019libritts} (target) with VoxCeleb2 \cite{Chung2018VoxCeleb2} (interference). The official split sizes are as follows: the training set contains 1,151 LibriTTS speakers and 5,900 VoxCeleb2 speakers with 149,681 utterances, corresponding to 250 hours of mixture data; the validation set contains 40 LibriTTS speakers and 94 VoxCeleb2 speakers with 4664 utterances, corresponding to 7.77 hours; and the test set contains 39 LibriTTS speakers and 118 VoxCeleb2 speakers with 3931 utterances, corresponding to 6.55 hours. The original mixing rule draws SNR uniformly from $[-5,5]$ dB. The dataset also provides a synthetic variant (\emph{Libri2Vox-syn}), made with speech generative models such as SALT \cite{Miaosynvox2} and SynVox2 \cite{Lv2023SALTDS}.
For evaluation, we construct test sets with 1, 2, and 3 interfering speakers by mixing the LibriTTS and VoxCeleb2 test partition. For mixtures with single interfering speaker, the SNR is sampled uniformly from $[-5,5]$ dB, following the setting of the original dataset \cite{Liu2024Libri2Vox}. For mixtures with 2 or 3 interfering speakers, the SNR range is adjusted to $[0,10]$ dB, in order to avoid energy overflow and clipping artifacts that may arise from multiple concurrent speakers. All test mixtures are with overlap ratio being 0.0, containing only real audio without synthetic interfering speakers.


\subsection{TSE Model Architecture} 
We employ a 2-layer 256-dimensional BLSTM-based network, followed by a fully connected network. that processes spectral features alongside speaker embeddings. The network takes 80-dimensional log-mel filterbank features as input, with speaker information provided by a pre-trained ECAPA-TDNN \cite{Desplanques2020ECAPA-TDNN} that remains frozen during training. The BLSTM processes temporal dependencies bidirectionally, generating time-frequency masks that isolate the target speech. Details about the model can be found in \cite{Liu2024Libri2Vox}.
Furthermore, CL strategy has demonstrated transferability across architectures, suggesting that once effective on a particular model, it can be readily applied to other separation networks as well~\cite{Liu2024Target}.

\subsection{Training Setup}
Optimization employs the Adam optimizer with an initial learning rate of 2e-4, combined with a warm-restart learning rate schedule. The warmup phase spans 5,000 steps with exponential increase to 1e-3, followed by Noam-style decay proportional to the inverse square root of the step count, ultimately decreasing to 1e-5 \cite{Vaswani2017Attention}. This schedule ensures stable initial learning while maintaining adaptability in later stages. 
Early stopping monitors validation loss with patience of 6 epochs for each stage, preventing overfitting while ensuring complete convergence. 
We evaluate performance using the SDR as the primary metric, and additionally report the improvement SDR (iSDR) relative to the input mixture, a measure widely adopted in previous studies \cite{2019SpeakerBeam, Xu2020SpEx}.

\section{Results}


\subsection{Analysis of single-factor vs. multi-factor}

To understand the relative contribution of different complexity factors and validate the effectiveness of our multi-factor approach, we conduct systematic comparisons between single-factor curricula and our proposed multi-factor strategy.

The baseline employs uniform random sampling without curriculum learning, where training examples are drawn with SNR uniformly sampled from $[0, 10]$~dB, number of interfering speakers from $\{1,2,3\}$, overlap ratio from $\{0,0.2,0.4\}$, and interference type from $\{\text{real}, \text{syn},\text{real/syn} \}$. This achieves iSDR values of 12.38~dB, 8.56~dB, and 7.16~dB for 2-speaker, 3-speaker, and 4-speaker mixtures respectively. The baseline-CL introduces a basic three-stage curriculum with fixed parameters, progressively increasing the number of speakers from 1 to 3 across stages while maintaining other factors constant, yielding marginal improvements to 12.58~dB, 8.78~dB, and 7.40~dB respectively.

Using baseline-CL as our reference point, we systematically evaluate single-factor curricula by modifying only one complexity factor while keeping others constant. Among single-factor curriculum strategies, the SNR-based curriculum achieves notable gains with 13.04~dB, 9.71~dB, and 8.62~dB by progressively adjusting SNR thresholds across stages.  The single-factor speaker count experiment fixes all stages to use only 1 interfering speaker, in contrast to the default curriculum learning that progressively increases from 1 to 3 speakers per stage, resulting in performance of 12.80 dB, 7.79 dB, and 6.73~dB. Training exclusively with single interfering speakers achieves better performance on 2-speaker mixtures compared to baseline-CL, but shows significant degradation on multi-speaker scenarios, indicating that incorporating multi-speaker cases during training introduces a trade-off that slightly costs 2-speaker performance while improving on more complex scenarios. The multi-factor strategy achieves the best overall performance with 13.22~dB, 10.08~dB, and 9.21~dB respectively. This represents substantial improvements of up to relatively 24.5\% over baseline in the 4-speaker case, with gains increasing as the number of interfering speakers grows, validating that coordinated multi-factor progression enables more effective learning than single-factor optimization.

\begin{table}[t]
\centering
\caption{Impact of single-factor and multi-factor CL. Each row represents a different curriculum configuration with varying factors. The curriculum stages column shows the progression of training difficulty across three stages, with parameters in the format $[\textit{num\_speakers}, \textit{snr\_low}, \textit{snr\_high}, \textit{overlap\_ratio}, \textit{inter\_source}]$. \textit{num\_speakers}: number of interfering speakers, [\textit{snr\_low}, \textit{snr\_high}]: SNR range in dB, \textit{overlap\_ratio}: fraction of temporal overlap, ranged from 0.0 to 0.4, \textit{inter\_source}: whether interference comes from real or synthetic (syn) speakers.}
\label{tab:ablation}
\scalebox{0.75}{
\begin{tabular}{l l c c c}
\toprule
\textbf{Exp.} & \textbf{Curriculum Stages} & \textbf{iSDR 2spk} & \textbf{iSDR 3spk} & \textbf{iSDR 4spk} \\
\midrule
baseline & random sampling & 12.38 & 8.56 & 7.16 \\
\midrule
baseline-CL & \begin{tabular}[c]{@{}l@{}}
{[}[1,0,10,0,real], \\
{[}2,0,10,0,real], \\
{[}3,0,10,0,real]]
\end{tabular} & 12.58 & 8.78 & 7.40 \\
\midrule
\makecell[l]{single factor:\\ \#snr} & \begin{tabular}[c]{@{}l@{}}
{[}[1,5,10,0,real], \\
{[}2,0,10,0,real], \\
{[}3,0,5,0,real]]
\end{tabular} & 13.04 & 9.71 & 8.62 \\
\midrule
\makecell[l]{single factor:\\ \#overlap ratio} & \begin{tabular}[c]{@{}l@{}}
{[}[1,0,10,0,real], \\
{[}2,0,10,0.2,real], \\
{[}3,0,10,0.4,real]]
\end{tabular} & 12.62 & 9.87 & 8.76 \\
\midrule
\makecell[l]{single factor:\\ \#inter\_source\_syn} & \begin{tabular}[c]{@{}l@{}}
{[}[1,0,10,0,syn], \\
{[}2,0,10,0,syn], \\
{[}3,0,10,0,syn]]
\end{tabular} & 11.22 & 8.43 & 9.19 \\
\midrule
\makecell[l]{single factor:\\ \#inter\_source\_real/syn} & \begin{tabular}[c]{@{}l@{}}
{[}[1,0,10,0,real/syn], \\
{[}2,0,10,0,real/syn], \\
{[}3,0,10,0,real/syn]]
\end{tabular} & 12.73 & 9.51 & 8.96 \\
\midrule
\makecell[l]{single factor:\\ \#speaker count} & \begin{tabular}[c]{@{}l@{}}
{[}[1,0,10,0,real], \\
{[}1,0,10,0,real], \\
{[}1,0,10,0,real]]
\end{tabular} & 12.80 & 7.79 & 6.73 \\
\midrule
multiple factors & \begin{tabular}[c]{@{}l@{}}
{[}[1,5,10,0,real/syn], \\
{[}2,0,10,0.2,real/syn], \\
{[}3,0,5,0.4,real/syn]]
\end{tabular} & \textbf{13.22} & \textbf{10.08} & \textbf{9.21} \\
\bottomrule
\end{tabular}
}
\end{table}

\subsection{TSE-Datamap Analysis}

Here we present the results and findings with TSE-Datamap. To determine the optimal ordering of difficulty regions for CL, we conduct experiments with all possible permutations of easy (E), ambiguous (A), and hard (H) examples. Each permutation represents a different strategy where training progresses through three stages with equal duration. The generation of each area follows the methodology described in Section 2.4.

\begin{table}[t]
\centering
\caption{Performance comparison of different stage orderings in curriculum learning. Each stage takes one third of total epochs. Baseline uses random sampling without curriculum.}
\label{tab:stage_ordering}
\begin{tabular}{l|ccc}
\hline
Order & iSDR 2spk & iSDR 3spk & iSDR 4spk \\
\hline
baseline & 12.38 & 8.56 & 7.16 \\
multiple factors & \textbf{13.22} & \textbf{10.08} & 9.21 \\
\hline
E / A / H      & 13.15 & 9.85 & \textbf{9.32} \\
E / H / A      & 12.93 & 9.63 & 9.18 \\
A / E / H      & 12.82 & 9.61 & 9.22 \\
A / H / E      & 12.96 & 9.72 & 9.17 \\
H / E / A      & 12.77 & 9.54 & \textbf{9.32} \\
H / A / E      & 12.90 & 9.63 & 9.10 \\
\midrule
E / A / H (forgetting) & 8.83 & 5.43 & 5.52  \\
\hline
\end{tabular}
\end{table}

The results of multiple TSE-Datamap variants are shown in Table~\ref{tab:stage_ordering}, along with the baseline and the multi-factor systems, which reports best performance in Table 1. The E / A / H schedule performs the best among all schedules, leading to absolute gains of 0.77, 1.29, and 2.16 dB over baseline, and outperform crafted multi-factor solution by 0.11 dB. The benefits grow with more interferers, which indicates the potential of TSE-Datamap against the increasing complexities.
Among easy-first strategies, E / A / H outperforms E / H / A, suggesting that ambiguous data before hard cases better calibrates decision boundaries. In contrast, orders placing hard examples earlier (A / E / H, A / H / E) or starting with them (H / E / A, H / A / E) generally underperform due to unstable early optimization. Overall, E / A / H provides the most effective progression, especially in challenging multi-speaker settings.

We also conduct an E / A / H (forgetting) experiment, where the data used for each stage is same as E / A / H, but each stage uses only its target region data without retaining previous stages. Such setup shows notable performance degradation, confirming that catastrophic forgetting occurs when earlier learned knowledge is not maintained during curriculum progression.

\subsection{Fixed-Quantity Ablation of Datamap Regions}
\begin{table}[t]
\centering
\caption{iSDR (dB) under fixed data quantity drawn from Datamap regions. All settings use 70\% of the training set for equal-quantity comparison. }
\label{tab:fixed_quantity_regions}
\begin{tabular}{lccc}
\toprule
\textbf{Setting} & \textbf{2spk} & \textbf{3spk} & \textbf{4spk} \\
\midrule
ambi70\% & \textbf{11.67} & \textbf{8.67}  & \textbf{8.61}  \\
baseline70\% & 11.37 & 8.19 & 7.17 \\
hard70\% & 10.10 & 6.75 & 6.99 \\
easy70\% & 9.71 & 6.50 & 6.40 \\
\bottomrule
\end{tabular}
\end{table}
To isolate the learning contribution of different TSE-Datamap regions from data quantity effects, we conduct controlled ablation experiments under fixed data quantity. We constrain all methods to utilize exactly 70\% of the total training data. The all70\% baseline samples uniformly from the entire training distribution, while easy70\%, ambi70\%, and hard70\% allocate 70\% of their data to their respective target regions. For all cases, the remaining 30\% of the data is distributed evenly between the other two regions.

As shown in Table~\ref{tab:fixed_quantity_regions}, under identical data quantity constraints, ambi70\% consistently outperforms the all70\% baseline across all mixture complexities. The easy70\% setting performs worst among the four approaches, which may due to these examples being rapidly handled, providing diminishing informative gradients as training progresses. In contrast, ambiguous examples maintain consistent informativeness throughout training due to their inherent learning difficulty, compelling the model for more robust decision boundaries that enhance generalization under equivalent data quantity.

\section{Conclusion}
In this study, we have presented TSE-Datamap, a framework for analyzing training dynamics in target speaker extraction by identifying easy, ambiguous, and hard regions based on confidence and variability metrics. Our multi-factor curriculum learning strategy jointly schedules SNR, speaker counts, overlap ratios, and synthetic/real proportions, achieving up to 24.5\% relative iSDR gains in multi-speaker scenarios. The Easy-Ambiguous-Hard ordering have been the most effective one among different schedules, validating that models benefit from establishing reliable decision boundaries before tackling complex cases.


\section{Acknowledgments}
This study is partially supported by MEXT KAKENHI Grants (24K21324) and JST, the establishment of university fellowships towards the creation of science technology innovation (JPMJFS2136).

\bibliographystyle{IEEEbib}
\bibliography{strings,refs}

\end{document}